\pdfoutput=1
\documentclass[conference]{IEEEtran}
\usepackage{cite}
\usepackage{amsmath,amssymb,amsfonts}
\usepackage{algorithmic}
\usepackage{graphicx}
\usepackage{textcomp}
\usepackage{xcolor}
\usepackage{fancyhdr}
\usepackage{outlines}

\usepackage[normalem]{ulem}
\usepackage{mathtools}
\usepackage[]{caption}
\usepackage{subfig}
\usepackage[]{algorithm2e}
\usepackage{multirow}
\usepackage{graphicx}
\usepackage{mathtools}
\usepackage{physics}
\usepackage{listings}
\usepackage[]{microtype}

\usepackage[hyphens]{url}

\def\BibTeX{{\rm B\kern-.05em{\sc i\kern-.025em b}\kern-.08em
    T\kern-.1667em\lower.7ex\hbox{E}\kern-.125emX}}

\newcommand{\mysizesmall}{0.35}

\newcommand{\amone}{{AM1}}
\newcommand{\amtwo}{{AM2}}
\newcommand{\pmgate}{{PM}}
\newcommand{\fm}{{FM}}
\newcommand{\is}{{IS}}
\newcommand{\gs}{{GS}}

\newcommand{\hide}[1] {}
\title{Architecting Noisy Intermediate-Scale \\ Trapped Ion Quantum Computers}

\author{\IEEEauthorblockN{Prakash Murali}
\IEEEauthorblockA{
Princeton University\\}
\and
\IEEEauthorblockN{Dripto M. Debroy}
\IEEEauthorblockA{Duke University}
\and
\IEEEauthorblockN{Kenneth R. Brown}
\IEEEauthorblockA{Duke University}
\and
\IEEEauthorblockN{Margaret Martonosi}
\IEEEauthorblockA{Princeton University}
}

\begin{document}
\maketitle
\pagestyle{plain}

\begin{abstract}

Trapped ions (TI) are a leading candidate for building Noisy Intermediate-Scale Quantum (NISQ) hardware. TI qubits have fundamental advantages over other technologies such as superconducting qubits, including high qubit quality, coherence and connectivity. However, current TI systems are small in size, with 5-20 qubits and typically use a single trap architecture which has fundamental scalability limitations. To progress towards the next major milestone of 50-100 qubit TI devices, a modular architecture termed the Quantum Charge Coupled Device (QCCD) has been proposed. In a QCCD-based TI device, small traps are connected through ion shuttling. While the basic hardware components for such devices have been demonstrated, building a 50-100 qubit system is challenging because of a wide range of design possibilities for trap sizing, communication topology and gate implementations and the need to match diverse application resource requirements. 

Towards realizing QCCD-based TI systems with 50-100 qubits, we perform an extensive application-driven architectural study  evaluating the key design choices of trap sizing, communication topology and operation implementation methods. To enable our study, we built a design toolflow which takes a QCCD architecture's parameters as input, along with a set of applications and realistic hardware performance models. Our toolflow maps the applications onto the target device and simulates their execution to compute metrics such as application run time, reliability and device noise rates.
Using six applications and several hardware design points, we show that trap sizing and communication topology choices can impact application reliability by up to three orders of magnitude. Microarchitectural gate implementation choices influence reliability by another order of magnitude. From these studies, we provide concrete recommendations to tune these choices to achieve highly reliable and performant application executions. With industry and academic efforts underway to build TI devices with 50-100 qubits, our insights have the potential to influence QC hardware in the near-future and accelerate the progress towards practical QC systems.

\end{abstract}

\section{Introduction}

\begin{figure}
    \centering
    \includegraphics[scale=0.4]{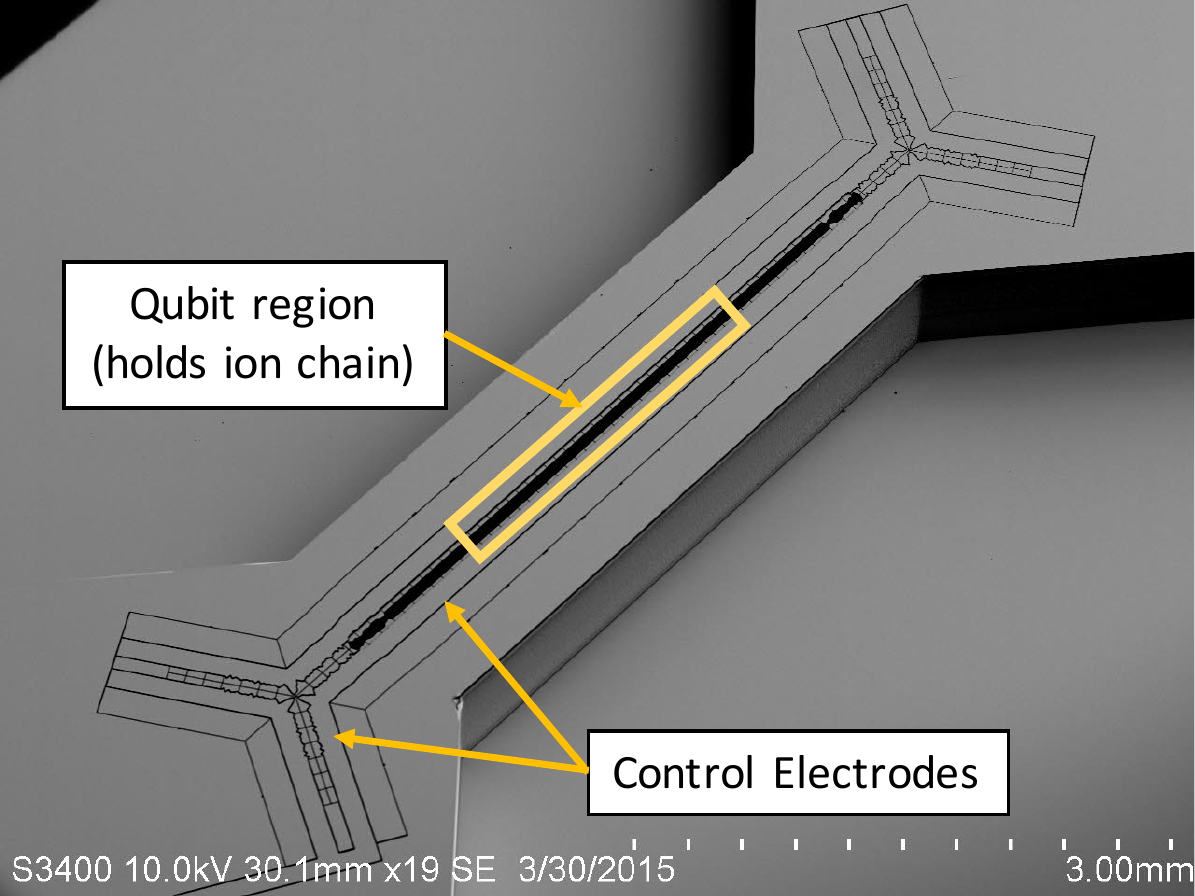}
    \caption{Scanning electron micrograph of the Sandia HOA2 trap. Figure adapted from \cite{Maunz2016}. A single trap houses all the ions. Control electrodes are used to load, remove and move ions. This architecture does not scale beyond 50-100 qubits because of gate implementation challenges in long ion chains.}
    \label{fig:sandia_trap}
\end{figure}
\begin{figure*}
    \centering
    \subfloat[5-qubit system with a single trap]{
    \includegraphics[scale=\mysizesmall]{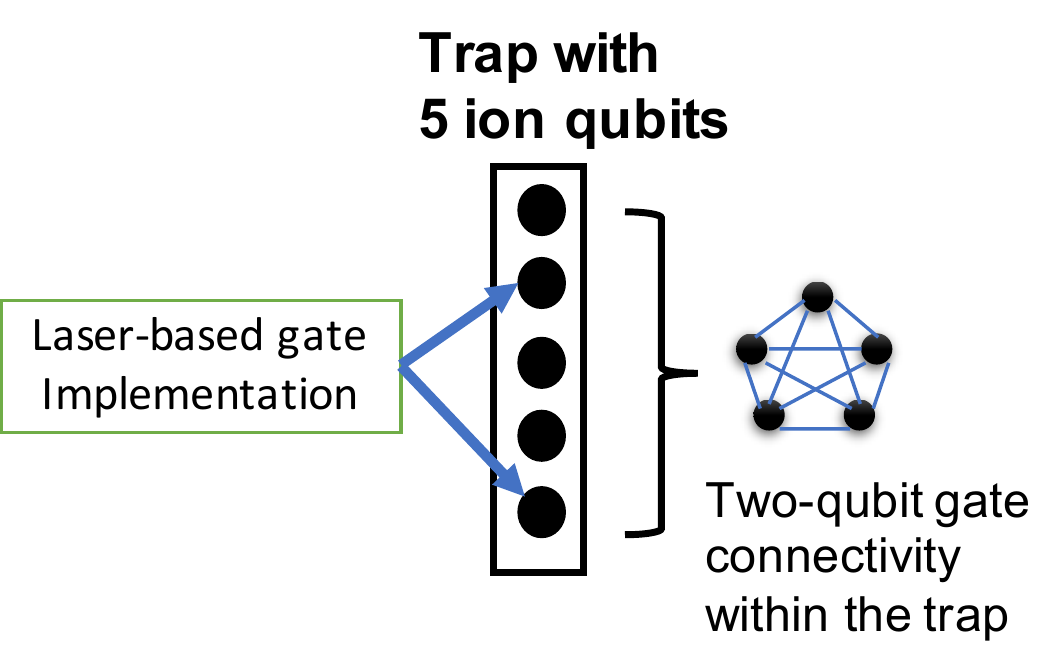}
    \label{fig:intro_one_trap}
    }
    \subfloat[Modular QCCD system]{
    \includegraphics[scale=\mysizesmall]{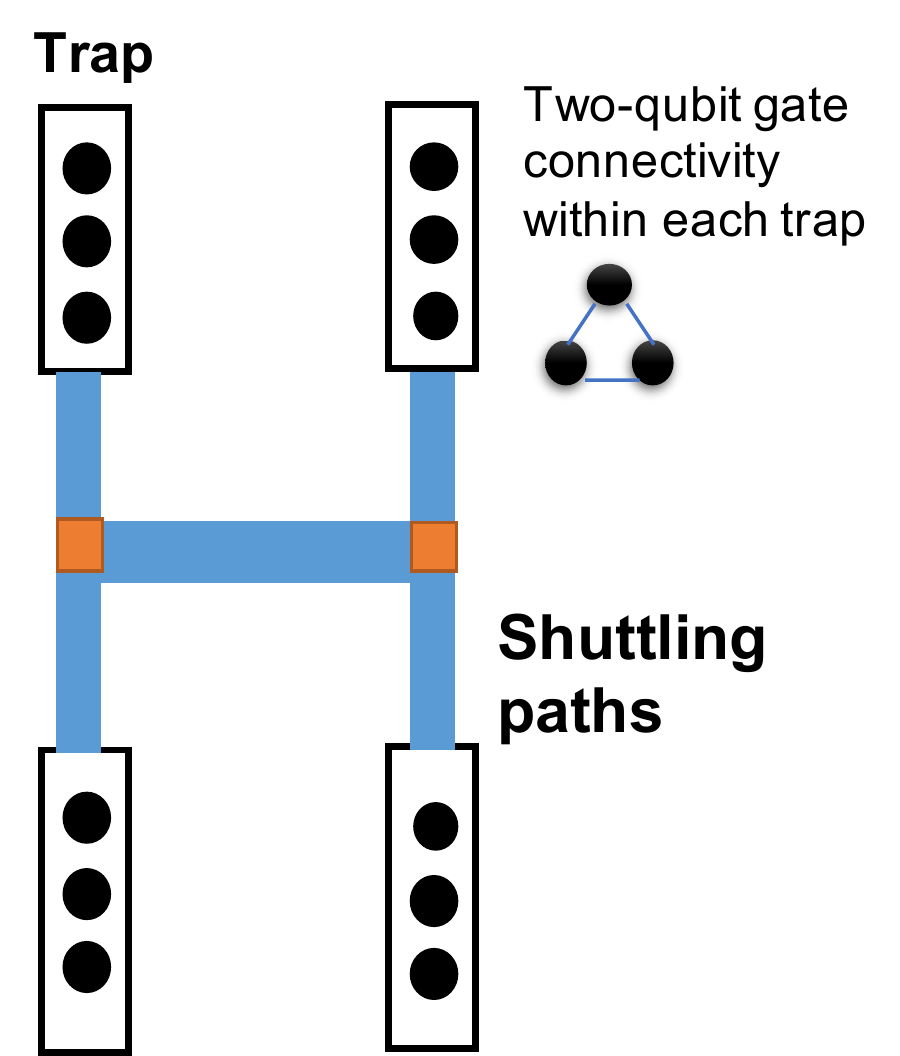}
    \label{fig:intro_qccd}
    }
    \subfloat[Example program IR]{
    \includegraphics[scale=\mysizesmall]{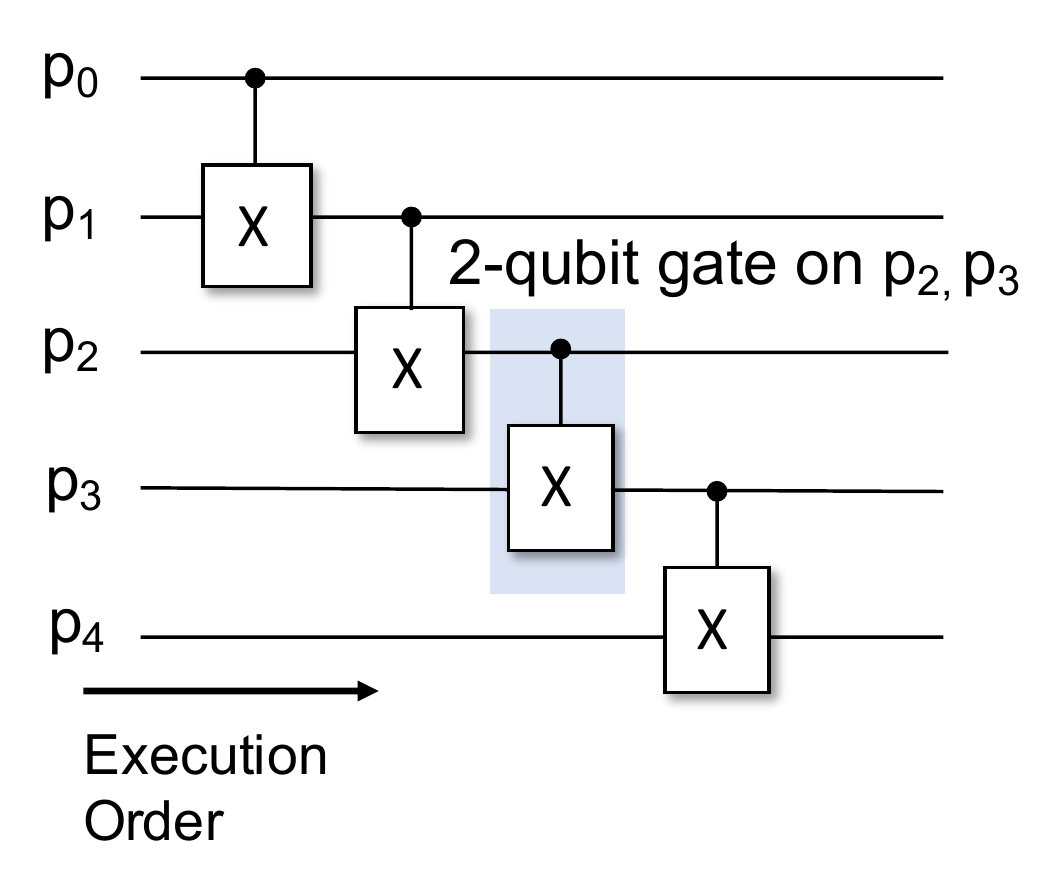}
    \label{fig:intro_ir}
    }
    \subfloat[Shuttling operation on $p_2$]{
    \includegraphics[scale=\mysizesmall]{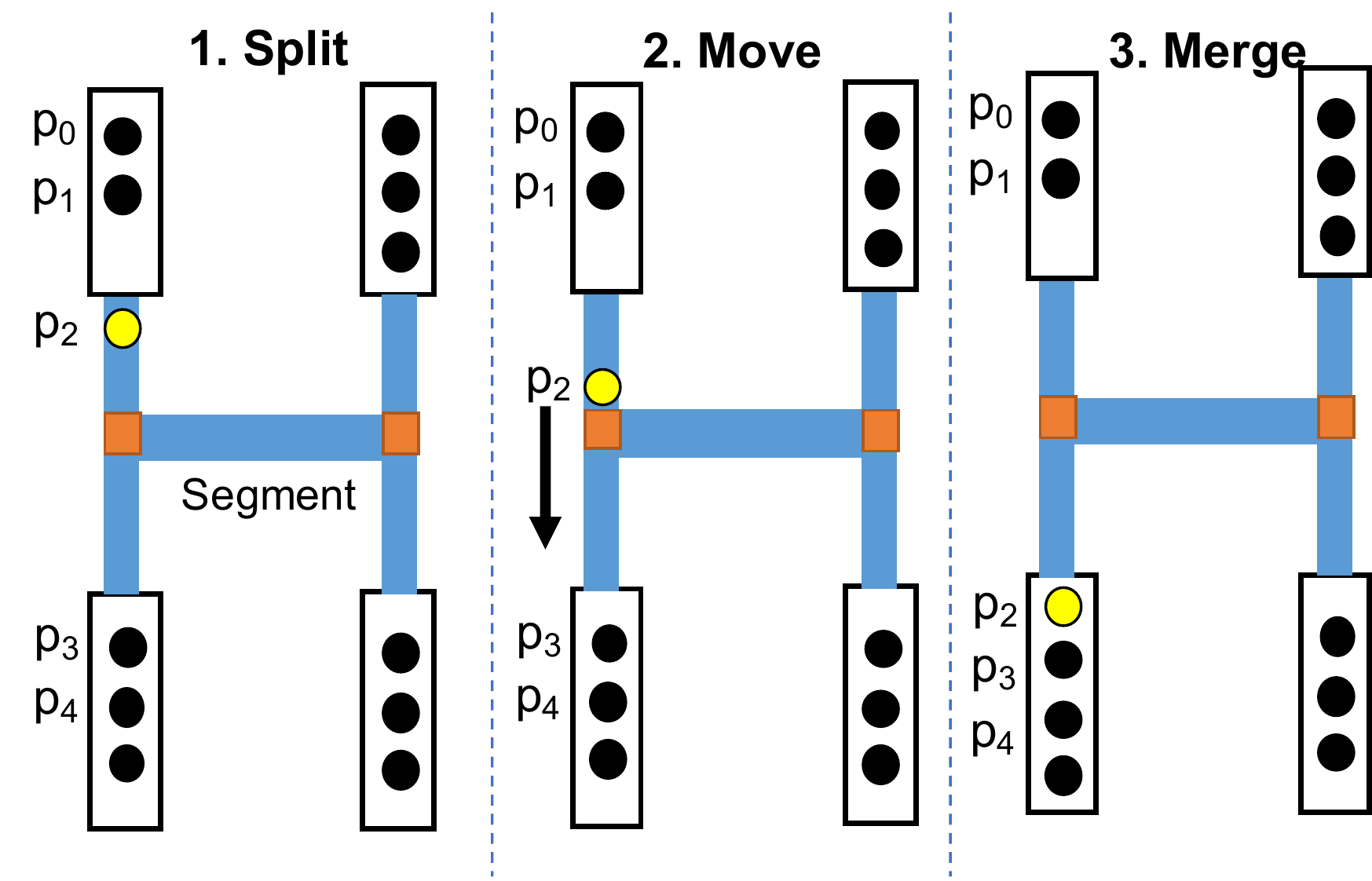}
    \label{fig:intro_shuttling}
    }
    \caption{(a) A 5-qubit TI system with a single trap. Each black circle represents a qubit. Two-qubit gates are performed by pulsing the desired pair of qubits with lasers, allowing a single trap to support full connectivity among the qubits. (b) A modular Quantum Charge Coupled Device (QCCD) with 4 traps. Each trap initially has 3 ions and a maximum capacity of 4 ions. The traps are interconnected through shuttling paths to move ions from one trap to another. The orange squares represent junctions where shuttling paths meet. (c) An example program intermediate representation (IR). For clarity, we show only two-qubit gates. Real program IR also includes single-qubit gates and qubit measurement operations. To execute the IR on the device in (a), each ion in the device can be used to represent one qubit from the IR, and gates can be executed using the laser controller. (d) To execute the IR on the device in (b), $p_0$, $p_1$ and $p_2$ are mapped onto one trap, and $p_3$ and $p_4$ are mapped onto another. The first two gates are executed within the top left trap. For the gate on $p_2$ and $p_3$, the qubits need to be co-located within the same trap, so $p_2$ is shuttled to the trap containing $p_3$ and the gate is performed inside the bottom left trap.} 
    \label{fig:intro}
\end{figure*}
\begin{figure}
    \centering
    \includegraphics[scale=0.4]{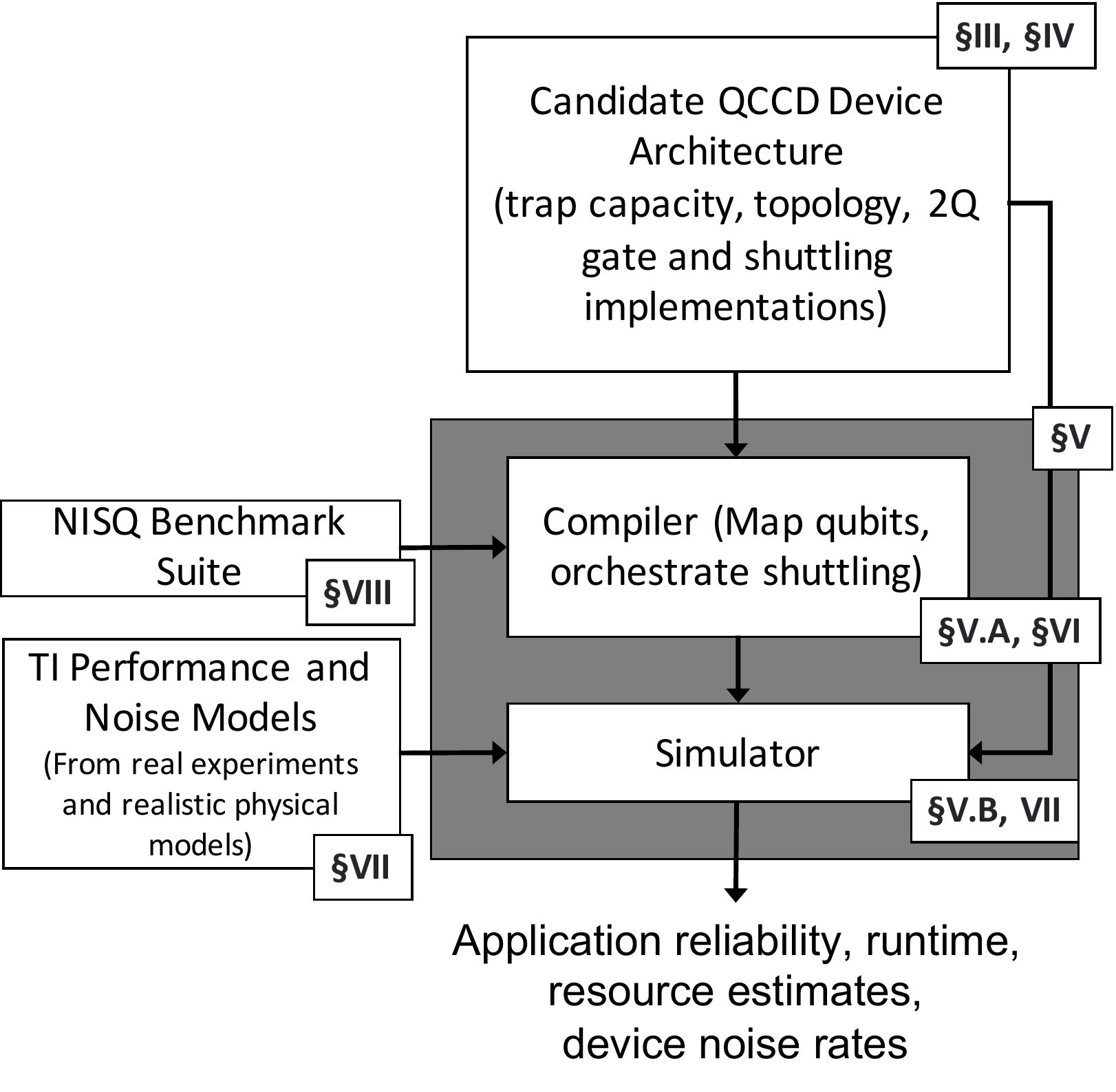}
    \caption{Our framework for evaluating a candidate QCCD-based TI system. Taking a candidate architecture, a set of NISQ applications, and realistic performance models as input, the toolflow computes application metrics like runtime and reliability (fidelity) and device metrics like heating rates.}
    \label{fig:our_framework}
\end{figure}
Quantum computing (QC) is an emerging paradigm which uses principles of quantum mechanics to manipulate information. In QC, information is represented using \emph{qubits} (quantum bits) and computations are performed using \emph{gates} (operations). Leveraging effects such as superposition, entanglement and interference, QC systems can efficiently explore exponentially large state spaces and compute solutions for certain classically-intractable problems. Practical applications of this paradigm are expected in the near future, particularly in the domains of computational quantum chemistry \cite{vqe1, vqe2}, machine learning \cite{quantum_nn1, quantum_ml1} and security \cite{supremacygoogle}.

QC hardware has progressed rapidly in recent years. Current leading qubit technologies are superconducting qubits \cite{superc1, superc2} and trapped ion (TI) qubits \cite{trappedion1, trappedion2, trappedion3}; other technologies also being pursued \cite{majorana2, majorana3, spinqubits}. Several superconducting qubit systems having up to 72 qubits have been built \cite{googlebristlecone, ibm50q, rigetti_transmons}. TI systems have also been built, with the current largest system having 11 qubits \cite{ionq11qubit}. 
All these systems have severe resource constraints, including low qubit counts and high operational noise, and therefore are called Noisy Intermediate-Scale Quantum (NISQ) systems.  In spite of these limitations, NISQ systems have the potential to demonstrate near-term QC applications \cite{supremacygoogle} especially if they are architected well and used in conjunction with efficient software toolflows. 



%


Trapped ion (TI) qubits are one the most promising technology candidates for building NISQ devices. Figure \ref{fig:sandia_trap} shows a real TI QC system. TI qubits are implemented using the energy states of an atomic ion such as Ca$^{+}$ or Yb$^{+}$. In a TI system, a set of ions are \emph{trapped} or confined in space using electromagnetic fields. As Figure \ref{fig:intro_one_trap} shows, the ions are arranged in the form of a linear chain, with each ion storing a single qubit. The states of the ions can be manipulated using lasers to implement gate-based computation. Current TI systems with 5-11 qubits have been used to demonstrate near-term applications and quantum error detection \cite{nam2019ground, ionq11qubit, linke2017fault, chiaverini2004realization}. Although they are smaller than superconducting systems (pursued by IBM, Google, Rigetti and others), they have fundamental advantages over other technologies, including defect-free identical qubits, very high coherence times \cite{wang2017single}, and dense qubit connectivity. Indeed, recent comparative studies show that TI systems perform better than superconducting systems of the same size \cite{linke2017experimental, triq}.



To scale up TI technology for near-term applications, academic and industry efforts are underway to build 50-100 qubit systems \cite{staq, honeywell, ionq}. Past this scale, architectures based on a single trap are infeasible because of difficulties in qubit control and gate implementation for long ion chains. Realizing these difficulties, a modular and scalable architecture called the Quantum Charge Coupled Device (QCCD) was proposed \cite{qccd1}. 
Figure \ref{fig:intro_qccd} shows an example. QCCD-based TI systems use multiple traps, with each trap having a small number of ions, allowing reliable gates and full connectivity within each trap. To interconnect traps, QCCD systems use \emph{ion shuttling}, where qubits are physically moved in order to allow communication between traps \cite{rowej2002transport, hensinger2006t, PhysRevLett.109.080501, shuttlingseparation_wineland1}. Figure \ref{fig:intro_ir} and \ref{fig:intro_shuttling} show an example shuttling operation. While several other scaling proposals exist in theory \cite{musiqc1, musiqc2, QLA, saqip}, all basic components required for QCCD systems have been developed and refined over the last decade \cite{shuttlingseparation_wineland1,PhysRevA.84.032314, Wright_2013, ionq11qubit,PhysRevLett.120.220501,leung2018robust,wan2020ion}, making it a very promising TI scaling path. Recently, Honeywell built the first QCCD system with 4 qubits \cite{pino2020demonstration} and shuttling-based systems are being pursued by other vendors also \cite{kaushal2019shuttlingbased}.

Although proof-of-concept QCCD systems have been demonstrated, building a large practical system is challenging. On the hardware and architecture side, designers face a wide range of design choices for trap capacity, device topology, gate implementation methods, and shuttling techniques. Currently there is little or no guidance on the performance and reliability tradeoffs of these choices. On the applications end, QC algorithms have widely different qubit and gate counts, error sensitivities, and communication patterns. If hardware is designed in isolation, without considering application characteristics, it will likely result in performance and reliability penalties that are too severe in the NISQ-regime. To enable practically useful hardware, computer architecture techniques must be applied to design TI devices that support application requirements well.

To this end, we perform an extensive architectural study of modular QCCD-based TI devices targeted for the 50-100 qubit range. Using a suite of NISQ applications, we evaluate a large space of design possibilities including key architectural choices and microarchitectural implementation methods. Figure \ref{fig:our_framework} shows our design tooflow for evaluating QCCD architectures. 
Our contributions include: 

First, while several works have focused on architecture for superconducting QC systems \cite{koen_bertels1, koen_bertels2, koen_bertels3, alimicro50, 1904.11590}, there has been less attention on TI systems although the technology is very promising. Our work performs the first architectural studies targeting systems with 50-100 qubits which are the next major milestone for TI systems. Our simulations emphasize the importance of optimizing the architecture --- across the hardware design space, application reliability varies up to five orders of magnitude depending on the choice of trap capacity, connectivity, and gate implementations. 

Second, our work provides concrete guidance for device designers as they architect larger systems. We find that having a capacity of 15-25 qubits per trap is ideal across applications and device topologies. This capacity range minimizes the impact of ion heating, laser beam instabilities, and motional energy hot spots across the device while still offering very good application performance. In addition, device topology must be co-designed for the needs of applications to achieve high reliability. For promising applications such as QAOA\cite{qaoa1, qaoa2}, linear device topologies work well and simplify hardware implementation. 

Third, our work provides insights on the best microarchitectural choices. We evaluate four entangling gate implementations and two methods for chain reordering and show that the most reliable implementations vary according to application characteristics i.e., microarchitecture must be re-configurable according to application requirements. 

\section{Background on Quantum Computing}
\label{sec:background}
\subsection{Principles of Quantum Computing}
{\noindent \textbf{Qubits:}} The building block of a QC system is a \emph{qubit} (quantum bit). Qubits have two basis states, $\ket{0}$ and $\ket{1}$. Using superposition, a qubit can be in a complex linear combination of the basis states, represented by $\alpha \ket{0} + \beta \ket{1}$, for $\alpha, \beta \in \mathbb{C}$. This allows an $n$-qubit system to potentially represent all $2^n$ basis states simultaneously, unlike a classical $n$-bit register which can be in exactly one of the $2^n$ states. 

{\noindent \textbf{Gates:}} To manipulate information, QC systems use \emph{gates} to modify the qubit amplitudes. Gates act on one or more qubits at a time. Similar to universal gates in classical computing, QC systems typically support a set of universal single-qubit and two-qubit gates. QC applications are expressed using these gate sets. To run a program, a sequence of gates is executed on a set of appropriately initialized qubits. The gates transform the qubit amplitudes, evolving the state space towards the desired output. To obtain classical output at the end of the algorithm, a qubit is measured, collapsing its state to either $\ket{0}$ or $\ket{1}$.



\subsection{Overview of Trapped Ion QC Systems}


{\noindent \textbf{Qubit Register (Ion Chain):}} In a TI quantum computer, information is stored in the internal states of ions which are trapped within an oscillatory potential \cite{ozeri2011trapped, wineland1998experimental}. DC electrodes on both ends of the trap provide a barrier along the axis of the trap, and a radio-frequency oscillating electric field fluctuates in the other two directions, causing the ions to be arranged as linear chain with even spacing. 

{\noindent \textbf{Qubit States:}} To store the $\ket{0}$ and $\ket{1}$ states required for QC, there are a wide variety of ion internal states, like hyperfine and Zeeman states, that can be chosen each having different strengths and weaknesses. The performance models used in our work assume qubits defined on hyperfine states, which is the standard choice in current devices. However, the insights from our work will also apply to other qubit states. 

{\noindent \textbf{Gate Implementation Using Lasers:}} Gates are implemented by exciting ions using lasers. Single qubit gates involve a single laser interacting with the desired ion, while two-qubit gates use multiple lasers, in order to excite the internal states of the ions and also the vibrational motions of the chain. Two-qubit gates use these joint oscillatory motions, also known as motional modes, as a bus to allow communication between internal states of distant ions \cite{trappedion1,PhysRevLett.82.1835,SorensenPRLMSgate1999}. The canonical two-qubit gate is the M\o{}lmer-S\o{}rensen  gate (MS), an entangling gate represented by a time evolution under an Ising type Hamiltonian, and is insensitive to the motional state of the ions. This motional state can cause issues with laser addressing of the ions, captured in the fidelity models we describe in Section \ref{sec:errormodel}.


{\noindent \textbf{Fidelity:}} In real QC systems, errors occur due to imperfect qubit control, errors in pulse implementation and external interference. \emph{Gate fidelity} refers to the quality of a gate measured using methods such as randomized benchmarking \cite{RB1}. For TI systems, gate fidelities higher than 99\% have been achieved in practice \cite{PhysRevLett.117.060504,ionq11qubit}. 

\section{Background on QCCD-based TI Systems}
\subsection{Challenges in Single Trap Architectures}
\label{sec:single_trap_scaling}
To motivate the design of QCCD-based systems we consider the challenges in scaling single trap systems to 50-100 qubits. First, within a single trap, the inter-ion spacing is determined by the balance between the trapping field and the Coulomb repulsion between the ions. When the ion count increases, the inter-ion spacing reduces, making it difficult to selectively pulse a qubit using laser controllers. 
Second, two-qubit gate implementation is also challenging. Within a trap, the ion-ion coupling strength for a pair of ions at distance $d$ scales in proportion to $1/d^\alpha$
with $\alpha$ ranging from 1 to 3 \cite{Zhang2017, leung2018entangling}. This increases the time required to perform an entangling gate on an arbitrary pair of qubits. Furthermore, the collective motional modes (vibrational modes) of the ion chain are used to mediate the two-qubit interaction. The density of modes increases with ion count, worsening the chance of crosstalk among modes and reducing gate fidelity\footnote{Ref. \cite{leung2018entangling} develops entangling gates on chains of 50 ions, but they see a considerable slowdown in two-qubit gate times.}. Put together, these challenges make it difficult to scale single-trap TI devices beyond tens of qubits. 

\subsection{Components of the QCCD Architecture}
QCCD devices overcome the challenges of single-trap systems using a modular design having a set of small ion chains, each in an individual trap. In Figure \ref{fig:intro_qccd}, the system has 12 ions, separated into 4 traps of size 3 each. By restricting capacity, this design achieves fast and high-fidelity two-qubit operations within each trap. To enable two-qubit gates across traps, QCCD uses ion shuttling to physically move ions from one trap to another prior to the entangling operation.

Figure \ref{fig:intro_shuttling} illustrates three steps involved in shuttling. First, the desired ion is \emph{split} from the source chain. To move this ion, shuttling paths are implemented as a set of segments connected by junctions. In Figure \ref{fig:intro_qccd}, the system has 5 segments (blue), connected using 2 junctions (orange). The split ion is moved from the trap through the segments and junctions to the desired trap. These \emph{move} operations also include any turns required at the junctions. Finally, the shuttled ion is \emph{merged} into the destination chain. Experimentally, these operations are implemented using time-varying waveforms on the control electrodes attached to the trap segments \cite{shuttlingseparation_wineland1}.

\section{Design Tradeoffs in QCCD-based TI Systems}
\subsection{Trap Capacity Choices}
Individual traps within a QCCD architecture are identical to a single-trap TI system, hence they face the same qubit addressing and gate implementation challenges if the number of ions in a single chain is too high. 
Therefore, having low trap capacity is beneficial to applications because it enables fast and reliable two-qubit gates within a trap. However, having low capacity is harmful because it sacrifices qubit connectivity, which is a key advantage of TI systems over other technologies. Satisfying an algorithm's two-qubit gate requirements with low trap capacity necessitates more shuttling, including more splits, moves, and merges. These operations increase execution time and reduce reliability. Further, shuttling operations introduce qubit motion via the trapping potentials and induce heating of the vibrational modes of the ion chain. This impacts qubit addressability using lasers and reduces the gate fidelities. 

Our work studies: \emph{How does trap sizing affect QCCD-based TI systems with 50-100 qubits? What sizes work well for NISQ applications and to what extent do application characteristics such as two-qubit gate patterns affect sizing?}

\subsection{Communcation Topology Choices}
\begin{figure}
    \centering
    \includegraphics[scale=0.35]{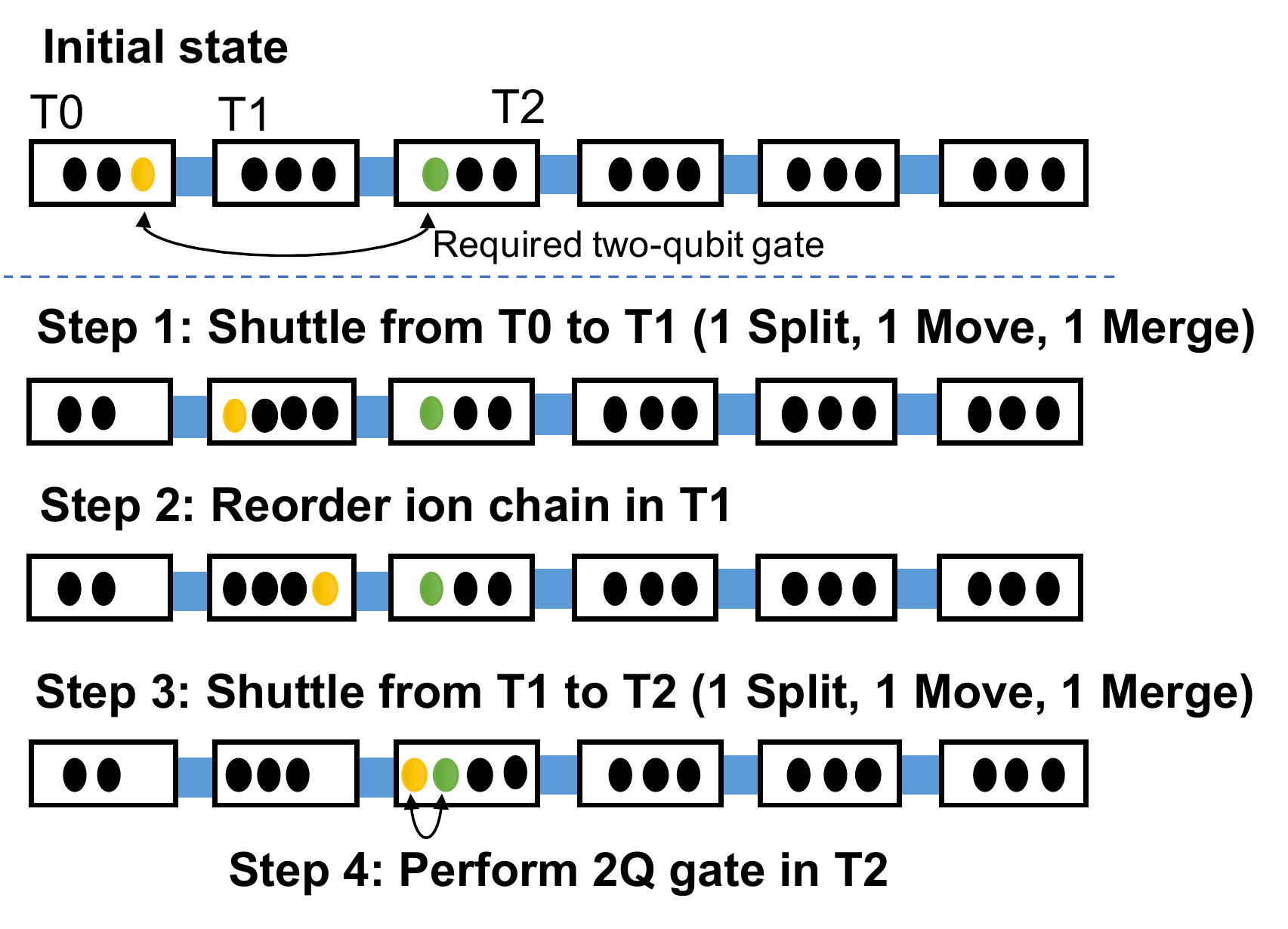}
    \caption{Shuttling in a QCCD-system which has linear device topology. Extra split and merge operations are required while moving ions through intermediate traps.}
    \label{fig:linear_top}
\end{figure}
%
QCCD systems have different topology options for orchestrating shuttling operations. 
To understand the tradeoffs, consider the linear topology shown in Figure \ref{fig:linear_top}. This topology is the easiest to build and imposes the minimum requirements on the number of required segments. Since there are no junctions, move operations are simplified. However, the linear topology restricts distant communication paths. To move an ion to a non-adjacent trap, several split and merge operations are required at intermediate traps. Splits and merges are more difficult compared to moves and can potentially impact applications. Additionally, split and merge operations require that the ion is positioned at the correct end of the chain. In our example, after the yellow ion is merged at the second trap, it needs to be repositioned at the right end of the second trap using a \emph{chain reordering} operation. These operations can also impact application metrics.
In contrast, grid topologies, such as Figure \ref{fig:intro_qccd} offer better communication paths at the expense of more hardware. In this particular 2x2 topology, shuttles do not encounter intermediate traps, and hence avoid the extra split, merge operations of the linear topology. However, grids require 3 and 4-way junction turns which are non-trivial compared to simple move operations through straight segments.

We ask: \emph{How much does QCCD device topology affect application reliability and performance? Are the overheads of extra split and merge operations in linear topologies prohibitive? 
What communication topologies can best support NISQ applications with 50-100 qubits?}

\subsection{Gate and Shuttling Implementation Choices}
{\noindent {\textbf{Two-qubit gates within a trap:}}} To implement two-qubit gates, the shared  motion of the ion chain can be harnessed in different ways. The two leading gate methods are based on amplitude modulation (AM) \cite{wu2018noise, am1_monroe, trout2018simulating} and frequency modulation (FM) \cite{leung2018robust,leung2018entangling} of the laser control pulses. We also consider a recent proposal based on phase modulation (PM) \cite{milne2018phase}.

To understand the impact of gate choices, consider a trap with $n$ ions, and say we wish to perform a gate between two ions that are separated by $d$ positions inside the trap. In Figure \ref{fig:intro_one_trap}, $n = 5$ and $d = 3$. With AM and PM gates, gate time linearly increases with $d$, i.e. gates between nearby ion pairs are faster than distant pairs assuming constant laser strength. This is a direct consequence of the weaker interaction strength between far away qubit pairs. On the other hand, for FM gates, duration is independent of $d$, but it increases linearly with $n$, i.e. for any qubit pair inside the trap, the gate time is constant, but as the gate times get longer as the chain does. These tradeoffs are not just in gate duration. Gate reliability worsens linearly with higher gate time and differs for AM, PM, and FM methods. Gate reliability also depends on heating rates, which are a function of the trap capacity and communication topology. Most importantly, since QC applications have diverse gate patterns, these tradeoffs are likely to play out differently across applications. It should be noted that none of these trends are fundamental. While there are methods to remove distance dependence for gate time and implementations with different scaling behavior, we consider the most commonly used pulse modulation techniques and base our studies on well-accepted experimental observations in the field. 

\begin{figure}
    \centering
    \includegraphics[scale=0.35]{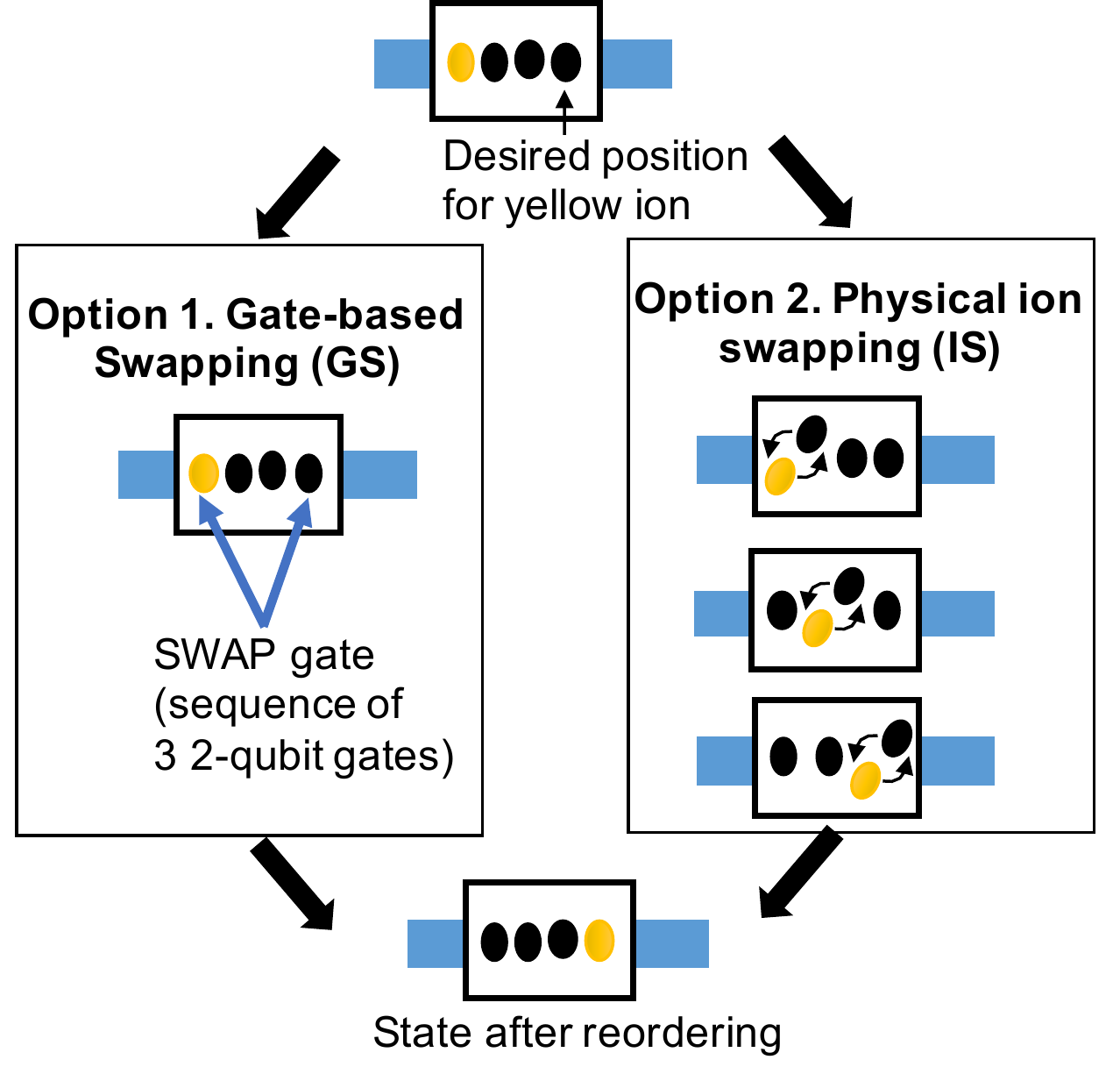}
    \caption{Choices for chain reordering. GS uses a SWAP gate (implemented with 3 MS two-qubit gates) to exchange quantum state of any arbitrary pair of ions within the trap. IS requires hop-by-hop physical swaps.}
    \label{fig:chain_reorder_choices}
\end{figure}

{\noindent {\textbf{Chain reordering within a trap:}}}
Another important microarchitectural choice is the method of chain reconfiguration. These operations position the ion at the correct end of the chain before a split operation (see Figure \ref{fig:linear_top}). The  two standard ways of performing reconfiguration: gate-based swapping and physical ion swapping, are shown in Figure \ref{fig:chain_reorder_choices}. In gate-based swapping (GS), a SWAP gate (implemented using 3 MS gates and some single qubit gates) is used to swap the quantum states of the desired ions. Hence, the performance and reliability of GS is directly influenced by the method for two-qubit gate implementation.
The second method, ion swapping (IS), physically swaps adjacent ions and was recently demonstrated \cite{ionswapping}. Each 1-hop IS exchange requires a split operation to isolate the two swapping ions, followed by the physical rotation of the two ions by 180 degrees (shown in Figure \ref{fig:chain_reorder_choices}), followed by a merge to reconstruct the chain (split and merge not shown). Similar to communication, split and merge operations for IS operations have performance and reliability overheads.
 
 We ask, \emph{What is the best method to implement two-qubit MS gates and chain reordering in near-term QCCD devices? Is the most reliable implementation different across applications? How can application characteristics be used to inform microarchitectural choices?}


\section{Design Toolflow: Overview}
To evaluate these design questions, we built the toolflow shown in Figure \ref{fig:our_framework}. Our framework takes a QCCD-based TI system design configuration as input, including trap sizes, connectivity, two-qubit gate implementation, and chain reordering method. It uses a set of NISQ application benchmarks to evaluate the candidate architecture. For accurate evaluation, our toolflow uses realistic performance models for individual components of the QCCD architecture, including real-system measurements reported in experimental works and realistic physical models. Our simulator uses these models to compute application-level metrics such as execution time, reliability, and operation counts along with device-level metrics such as trap heating rates. 

\subsection{Compiler for QCCD-based TI systems}
To evaluate a range of architectures, we require application executions that are optimized for each target architecture, ideally through an automated compiler toolflow. Current QC compilers \cite{qiskit, cirq, quilc, triq} do not support QCCD-based TI systems, so we built a backend compiler which maps and optimizes applications for QCCD systems. The input to the compiler is an application intermediate representation (IR) consisting of a gate sequence with data (qubit) dependencies among gates. Such IR can be obtained from the language frontends of QC compilers like IBM Qiskit \cite{qiskit, qiskit-terra} or ScaffCC \cite{scaffcc1, scaffcc2}. Using the IR, our compiler first maps the program qubits onto distinct hardware qubits using heuristic techniques which aim to reduce communication. Next, we route shuttling operations through the shortest paths in the hardware and automatically insert the necessary chain reordering operations. Since multiple shuttles are allowed to execute in parallel on QCCD devices, we implement strategies to avoid congestion at junctions and avoid deadlocks while routing parallel shuttles. The output of our compiler is an executable with primitive QCCD instructions.  

\subsection{Simulator using Realistic Performance Models}
\label{simulator_dicussion}
Next, we built a simulator to run the applications on the candidate architecture. The inputs to the simulator are the compiled executable, the target QCCD device and physical performance models for QCCD hardware. The goal of the simulator is to estimate application run time, reliability, and device-level metrics such as trap heating rates. 

To measure application run time, our simulator considers realistic gate performance models, shuttling time models and parallelism constraints in QCCD systems. The gate and shuttling performance models are derived from real device characterization studies, and allow us to accurately model the performance of all primitive operations in the QCCD architecture (Section \ref{sec:errormodel}). In TI systems, gates within a single trap typically execute serially \cite{ionq11qubit, triq}. But, independent ion shuttles can run in parallel with each other, and in parallel with gates in other traps. Considering these constraints, the simulator walks through the instructions in the compiled executable and schedules their execution on the device. The simulation begins with each qubit laid out acoording to the initial qubit layout specified by the executable. For shuttling operations, the simulator moves ion from one trap to another as specified by the executable. For each instruction the simulator tracks start and finish times, allowing it to estimate total application runtime at the end of the program.

To measure application reliability, we ideally require a quantum noise simulator. While such noise simulators have been developed \cite{qiskit_aer}, their compute requirements scale exponentially with qubit count and are intractable beyond 50-60 qubits. Moreover, current simulators are specific to superconducting qubits and do not include QCCD system models. Hence, we build a custom simulator for QCCD systems. Our simulator uses realistic physical models and estimates from real-system experiments to model gate fidelity and trap heating rates from operational and background noise sources.

The simulation starts with each chain in a zero motional mode energy state. When shuttling operations are executed, the motional energy of the ion chains increase (the ions vibrate more because energy is added to the system to move them). The simulator tracks these energy changes using estimates from a physical model. For each gate, the simulator computes the fidelity using a model which includes errors from chain temperature and background heating. To measure application reliability (fidelity), the simulator computes the product of fidelities for each operation in the program. This model closely approximates real executions and has been experimentally validated on current TI systems \cite{linke2017experimental, ionq11qubit}, on superconducting systems \cite{supremacygoogle}, and used in prior works \cite{asplos, rodvanmeter1}.


\section{Optimizations in Our Compiler}
The input IR to the compiler has a set of single and two-qubit gates, as well as measurement operations. Unlike classical compilation, QC IR does not have control dependencies. It is standard practice to fully unroll all loops and inline all functions; the full instruction sequence is known at compile time \cite{scaffold, asplos, triq, qiskit, cirq}.

Our compiler first maps the program qubits in the IR onto hardware qubits. For example, for the program in Figure \ref{fig:intro_ir}, $\{p_0, p_1, p_2\}$ can be mapped to the first trap, and $\{p_3, p_4\}$ can be mapped to the second trap. The choice of mapping influences the amount of communication. To reduce communication, we use a greedy mapping heuristic adapted from prior work \cite{cgo18, asplos}. Our heuristic orders the program qubits according to the sequence in which they are used by the application. It maps each qubit to a trap, co-locating qubits according to trap capacity constraints specified by the architecture. To leave enough buffer space for incoming shuttles, the heuristic ensures that traps are not completely filled (in our experiments, we leave room for 2 incoming ions per trap).

Next, to schedule gates, our compiler uses the earliest ready gate first heuristic \cite{heckey1}. Single-qubit gates on an ion do not need communication and can be scheduled on the trap which holds the ion. For two-qubit gates, the compiler inserts a series of split, move, and merge operations to co-locate the ions if required. To minimize communication, the compiler determines the shortest shuttling path using the device topology. Chain reordering operations are inserted automatically according to the method supported by the target device. For parallel shuttle routing, the compiler avoids deadlocks and manages congestion by leveraging full knowledge of the program instructions and device topology to allocate resources and break dependency cycles. When two shuttles need to access the same segments, we use heuristics to schedule the individual split, move, and merge operations such that no two ion shuttles occupy the same segment at the same time and prioritize earlier gates. To manage congestion at junctions, wait operations are inserted to delay ion shuttles at the intersection when another shuttle is passing or turning through the junction.

\section{Simulation Framework: Performance and Fidelity Models}
\label{sec:errormodel}
In order to effectively study architectural and microarchitectural design, we must have a well motivated model of the physical behavior of QCCD systems. First, we present performance models for two-qubit gate implementations. Next, we model shuttling performance and trap heating rates. Finally, the gate and shuttling models are combined according to a well-known model to compute the gate fidelity.

\subsection{Gate Time Model}
\label{sec:gateduration}
The entangling gate we consider is the M\o{}lmer-S\o{}rensen gate, which is the canonical two-qubit gate on TI systems. This gate creates entanglement between distant ions in a chain by mediating the interaction through the motional mode of the chain \cite{SorensenPRLMSgate1999, PhysRevLett.82.1835}. Other popular QC gates such as Controlled NOT are implemented using the MS gate as a low-level primitive \cite{MaslovCircuitCompIT2017,triq}. Our work considers three implementation options for the MS gate. The options differ by the laser parameter that is modulated to ensure robust performance on all motional modes. The first two techniques implement laser pulses using Amplitude Modulation (AM)  \cite{wu2018noise, trout2018simulating}, the third technique uses Phase Modulation \cite{milne2018phase}, and the last technique uses Frequency Modulation (FM)  \cite{leung2018robust, leung2018entangling}. These gates are standard and implemented in current TI systems \cite{ionq11qubit, blumel2019power, ball2020software}.

{\noindent \textbf{AM Gates:}} For the AM gates, the operation time of the gate is linearly proportional to the distance between the involved qubits. For clarity we refer to the gate implementation in \cite{wu2018noise} as AM1, and the gate implementation from \cite{trout2018simulating} as AM2. AM1 gates are slightly more robust to some sources of noise outside of the scope of this study, but as a result are slightly slower overall. Their gate times are well described by the function 
$$\tau_{AM1}\left(d\right) = 100*d - 22,$$
where $d$ is the number of ions between the two ions that are being entangled (all times are reported in $\mu s$). For the less robust but faster gates AM2 gates \cite{trout2018simulating}, we use
$$\tau_{AM2}\left(d\right) = 38*d + 10.$$

{\noindent \textbf{PM Gates:}} Similar to AM gates, the operation time of PM gates can be approximated as being linearly related to the distance between the involved qubits. From \cite{milne2018phase} we get a scaling of
$$\tau_{PM}\left(d\right) = 5*d + 160.$$
These gates have a much weaker dependence on distance than their AM counterparts, but are slower for nearby qubits.

{\noindent \textbf{FM Gates:}} FM gates have an operation time which is independent of the distance between the two ions, but are instead proportional to the total number of ions in the chain. We assume a gate time of $100\mu s$ for all chains below $12$ ions, as extremely fast gates are somewhat sensitive to noise, and then increase linearly from there according to the times given in \cite{leung2018robust}. From this we get an equation for gate time of 
$$\tau_{FM}\left(N\right) = max(13.33*N - 54, 100),$$
where $N$ is the number of ions in the chain.

\subsection{Shuttling Model}
{\noindent \textbf{Shuttling Time:}} During shuttling, the electrodes in the trap have their voltages modified split the chain and move the ion of interest slowly to the next segment. This slow motion is essential for minimizing the amount of heating present during the operation, but there still is some heating which is unavoidable.
In Table \ref{tab:shuttling} we give the times for the various shuttling operations, obtained from real characterization experiments \cite{gutierrez2019transversality}. These operations allow us to move ions between chains as needed in order to generate more complex entangled states and measure qubits, while still honoring trap capacity restrictions.

\begin{table}[h]
    \centering
    \begin{tabular}{c|c}
         Operation & Time \\ \hline
         Move ion through one segment & $5\mu s$\\
         Splitting operation on a chain & $80\mu s$\\
         Merging an ion with a chain & $80\mu s$\\
         Crossing Y-junction & $100\mu s$\\
         Crossing X-junction & $120\mu s$
    \end{tabular}
    \caption{Operation times for each shuttling operation, obtained from experimental demonstrations summarized in \cite{gutierrez2019transversality}.}
    \label{tab:shuttling}
\end{table}

{\noindent \textbf{Heating Model:}} When shuttling a qubit, motion is being introduced to the system via the trapping potentials, and this can cause additional heating of the motional modes of the chain. While our entangling gates do not need the chain to be in a particular motional mode, higher energy states have more vibration, making ideal laser addressing difficult. This leads to a penalty in gate fidelity as energy is added to the system.

In our heating model, every chain is thought of as a quantum oscillator with discrete quantized energy levels. We initialize all chains in the zero energy state of this system and add energy to the system in fractions of the energy difference between these energy levels, known as a quanta. When a chain is split, the energy of the chain is split proportionally to the number of ions in each sub-chain, such that conservation of energy is obeyed. Each sub-chain then gains $k_1$ quanta of motional energy. Similarly when two chains are merged, the resulting chain has energy equal to the sum of the two chains which are being merged, along with an additional $k_1$ quanta to account for the energy needed to stop the chains and prevent collisions. Lastly, when an ion is being shuttled, it picks up $k_2$ quanta of energy per segment it shuttles over, to account for slight imperfections in the shuttling potential, along with fact that the very act of shuttling requires the ion to increase in energy. This model comes from the intuition that the heating is strongest at points where the ion is experiencing higher accelerations, and that adiabatic shuttling has been shown to have high fidelities in practice \cite{huber2008transport,Eble:10,PhysRevA.84.032314}. 

In Honeywell's 4-qubit QCCD system, the average heating rate per shuttling operation was measured to be less than $2$ quantas per second \cite{pino2020demonstration}. Since further improvement will be necessary for realizing 50-100 qubit systems, we assume an order of magnitude lower heating rates and use $k_1=0.1$ and $k_2=0.01$.

\subsection{Gate Fidelity Model}
When assuming ideal addressing, the fidelity of the MS gate is independent of the motional mode. In practice, thermal motions from higher motional modes reduce the fidelity of the gate. Additionally, if background heating from the electric fields in the trap occurs during the gate, that gate will fail as the MS gate relies on a constant motional mode during application. These two effects lead to a gate fidelity $F$ defined as:
\begin{align}
F = 1 - \Gamma\tau - A\left(2\bar{n} + 1\right), 
\label{eq:fidelity}
\end{align}

where $\Gamma$ is the background heating rate of the trap, $\tau$ is the gate duration defined in Section \ref{sec:gateduration}, $A \propto\frac{N}{ln (N)}$ is a scaling factor on the second term which represents the thermal laser beam instabilities (thermal motion of the laser beams perpendicular to the ion chain), and $\bar{n}$ is the motional mode of the chain (vibrational energy), in units of motional quanta \cite{wu2018noise}. In other words, fidelity decreases at higher gate durations because of background heating. Fidelity also decreases when the motional energy of the chain increases from shuttling operations.
\section{Experimental Setup}
\begin{table}[h]
    \centering
    \begin{tabular}{c|c|c|c}
    Application & Qubits & Two-qubit Gates & Communication Pattern\\ \hline
    Supremacy & 64 & 560 & Nearest neighbor gates\\
    QAOA & 64 & 1260 & Nearest neighbor gates\\
    SquareRoot & 78 & 1028 & Short and long-range gates\\ 
    QFT & 64 & 4032 & All distances (64*63 gates) \\
    Adder & 64 & 545 & Short range gates \\
    BV & 64 & 64 & Short and long-range gates
\end{tabular}
    \caption{Applications used in our study. 
    }
    \label{tab:applications}
\end{table}
\subsection{Applications}
Table \ref{tab:applications} lists the six applications used in our study. 
Google's recent supremacy demonstration used a circuit with 53 qubits and 430 two-qubit gates on real superconducting hardware \cite{supremacygoogle}. Using this as a baseline capability for 50-100 qubit NISQ systems, we selected applications with 60-80 qubits and 500-4000 two-qubit gates. 

The quantum supremacy benchmark is designed to demonstrate a classically intractable computation on a near-term QC system \cite{boxio, supremacygoogle}. Quantum Approximate Optimization Algorithm (QAOA) is an important optimization algorithm with near-term applications \cite{qaoa1, qaoa2, ti_qaoa_expt}. We use the hardware efficient ansatz for QAOA described in 
\cite{Moll_2018}. SquareRoot is an implementation of Grover's search algorithm \cite{grover}. Quantum Fourier Transform (QFT) \cite{bernsteinvazirani} and Adder are important QC kernels. Bernstein-Vazirani (BV) algorithm has been used to characterize current trapped ion systems \cite{ionq11qubit}.

We obtained the IR for SquareRoot and QFT from ScaffCC \cite{scaffcc1, scaffcc2}, Supremacy from Google Cirq \cite{cirq}, and QAOA, BV and Adder from \cite{teague}. Our backend compiler supports an OpenQASM interface \cite{openqasm1} which allows us to easily interface with high-level language frontends like Cirq and ScaffCC.

\subsection{Device Configurations}
QCCD systems are designed to operate in the regime of 50-200 qubits. Beyond that optical interconnects and other scaling techniques are required to build very large systems with thousands of qubits \cite{musiqc1, musiqc2}. We evaluate architectures with 50-200 qubits and consider individual trap capacities in the range of \textbf{15-35 ions per trap}. To explore communication topologies, we use two device topologies: \textbf{L6}, a device similar to Figure \ref{fig:linear_top} with 6 traps connected in a linear fashion (this is the topology of Honeywell's QCCD system \cite{pino2020demonstration}), and \textbf{G2x3} a grid device similar to Figure \ref{fig:intro_qccd} with 6 traps arranged in two rows and three columns \cite{qccd1}. To test gate implementations, we consider 4 variants of the MS gate: \textbf{\amone}, \textbf{\amtwo}, \textbf{\pmgate}, and \textbf{\fm}. We also test two variants of chain reordering: \textbf{\gs} and \textbf{\is}. 


All compilation and simulation experiments are run on an Intel Skylake processor (2.6GHz, 12GB RAM) using Python 3.7.
\section{Architectural Design Exploration}
\label{sec:arch_design}
\begin{figure*}[t]
    \centering
    \includegraphics[trim = 0 0 0 0, width = \linewidth]{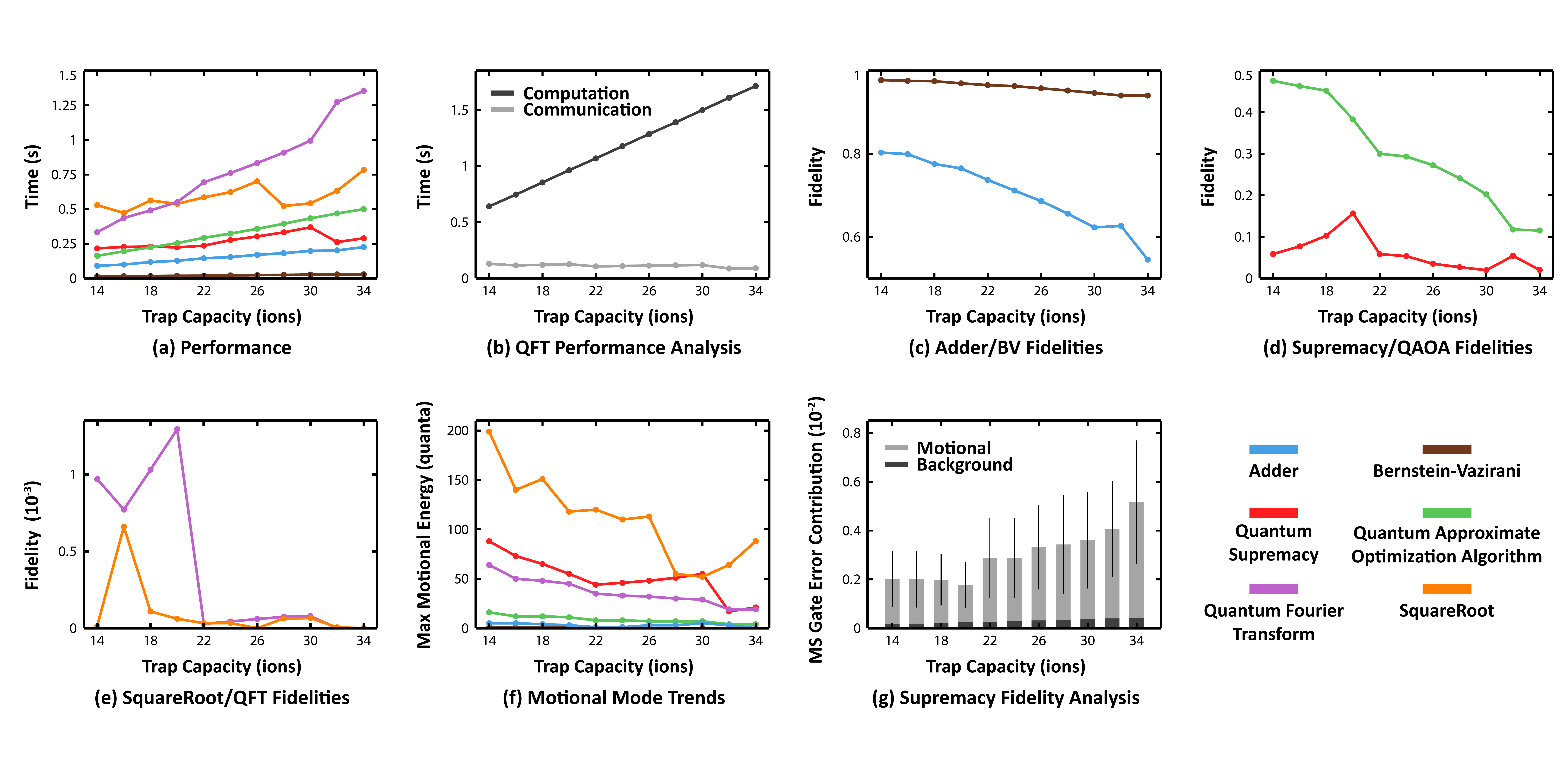}
\caption{\textbf{Trap Sizing Choices:} Experiments use L6 device, with FM two-qubit gates and GS chain reordering. Capacity denotes the maximum number of ions in an individual trap. (a) shows application runtime (lower is better). Runtime depends on trap capacity, but is also influenced by application characteristics. (b) shows the trends of computation and communication time for QFT. Communication time decreases with high trap capacity, while computation time increases because of higher gate time in large traps. (c-e) show application fidelity (product of gate fidelities, higher is better). \textbf{Application fidelity varies dramatically based on individual trap capacity}. 15-25 ions per trap works well across applications, with severe fidelity degradation beyond 35 ions. (f) shows maximum motional mode energy across the device (unwanted vibrational energy in ion chains, lower is better). Motional mode energy decreases at higher capacity because of reduced communication. (g) analyzes the contribution of background heating and motional mode energy to two-qubit gate error rate (error rate is $1-$gate fidelity, lower is better). Motional mode energy is the major contributor to heating error. The trend is explained in Section \ref{sec:trap_capacity_choices}.
\label{fig:trap_capacity_choices}
}
\end{figure*}

\begin{figure*}[h]
    \centering
    \label{fig:comm_choices}
    \includegraphics[trim = 0 0 0 0, width = \linewidth]{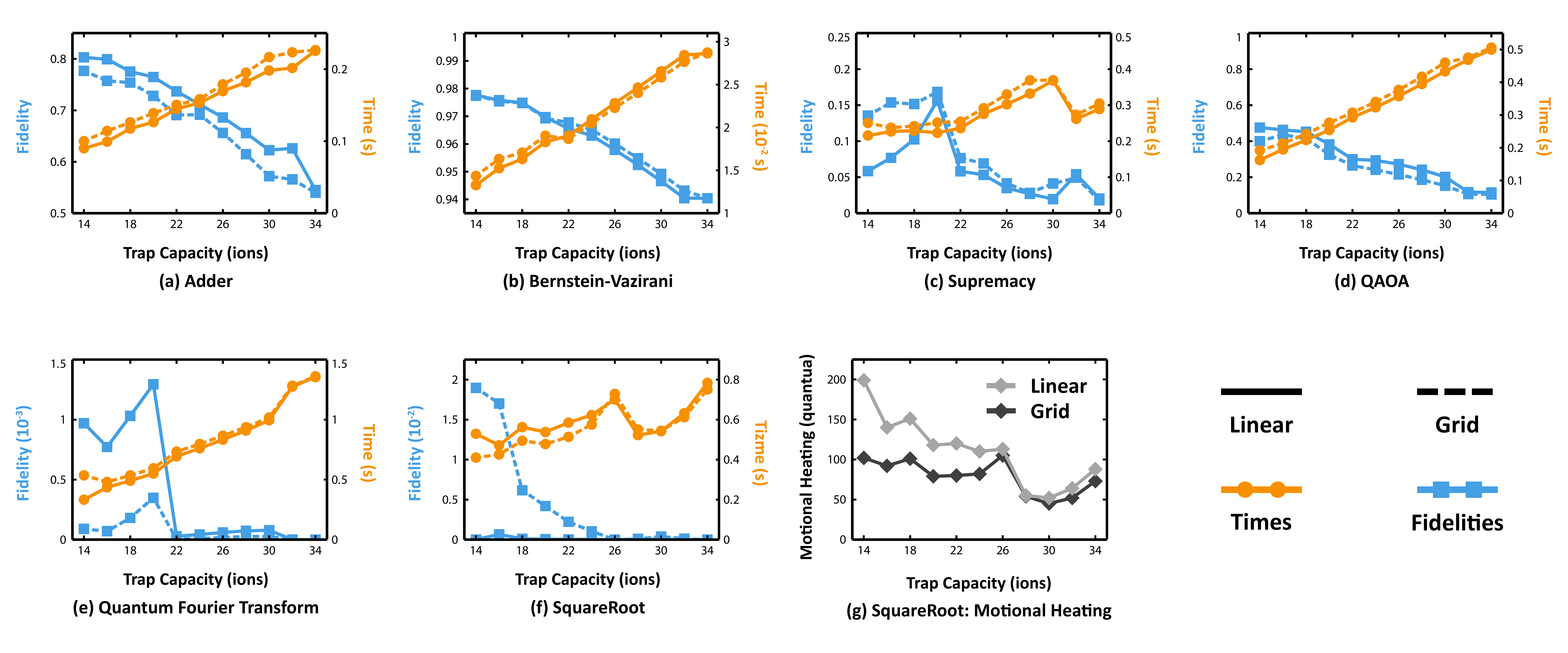}
\caption{\textbf{Communication topology choices:} Figure compares two topologies: L6 and G2x3. Experiments used FM two-qubit gates with GS reordering. (a)-(f) show application runtime (lower is better). Topology affects performance, depending on application characteristics. (e-h) show application fidelity (higher is better). \textbf{Application fidelity is significantly impacted by communication topology. When application and device topology are well-matched, fidelity is boosted by up to 3 orders of magnitude.} (g) shows motional mode energy for SquareRoot (lower is better). Grid topology offers high fidelity for this application because it reduces communication operations, and hence has lower motional mode energy.}
\label{fig:comm_choices}
\end{figure*}
\subsection{Trap Capacity Choices}
\label{sec:trap_capacity_choices}
Figure \ref{fig:trap_capacity_choices} shows the effect of trap sizing on application and device-level metrics. Figure \ref{fig:trap_capacity_choices}a shows the execution time (performance) for the six applications (lower is better). For SquareRoot, Supremacy and BV, the performance is relatively stable with increasing capacity. This arises because of relative amounts of compute and communication and the different scaling trends for these components. As trap capacity increases, the amount of communication drops. 
However, the gate time increases because longer duration is necessary to perform entangling gates in large traps (see FM gate scaling in Section \ref{sec:gateduration}). Hence, the overall time remains relatively constant irrespective of trap size.
However, Figure \ref{fig:trap_capacity_choices}b analyses the computation and communication performance for QFT. In this case, computation time is the dominant factor and the total time increases with trap size. Therefore, while it is generally believed that the shuttling time will be a major performance bottleneck for QCCD systems \cite{mit_ti_survey, PhysRevLett.120.220501}, our work shows that computation and communication performance depend on application characteristics as well as device architecture. 

Figure \ref{fig:trap_capacity_choices}c-\ref{fig:trap_capacity_choices}e show the fidelity of six applications (higher is better). For BV, Adder and QAOA,  fidelity is high even at very low trap capacity because of their low communication requirements. For Supremacy, SquareRoot and QFT, fidelity is low at small trap capacity ($<$ 15 ions), attains a maximum thereafter and drops significantly when the trap capacity is 30 or more. For Supremacy, the best fidelity is $15\times$ higher than the worst, showing the importance of optimizing trap sizing. To analyze the trend, Figure \ref{fig:trap_capacity_choices}f shows the maximum motional mode across the traps in the device (the motional mode quantifies unwanted energy accumulated in an ion chain, higher is worse). The motional energy is high at small capacity because more communication operations are required. Each shuttling operation adds energy to the ion chains, increasing heating, worsening qubit addressability and gate fidelity. Since heating rates reduce with increasing trap capacity, why does gate fidelity worsen at higher capacity?

Figure \ref{fig:trap_capacity_choices}g analyses the contribution of background heating and motional mode energy towards two-qubit gate errors for Supremacy (see equation \ref{eq:fidelity}). Gate error is dominated by the motional mode error, with only a negligible contribution from background heating. Surprisingly, even though the motional mode energies reduce at larger trap capacity, the thermal contribution to gate error increases with capacity --- the error rate increases by $3\times$ for a capacity of 35 ions, compared to 20 ions. This is for two reasons: First, laser beam instabilities increase with trap capacity (captured by the second term in equation \ref{eq:fidelity}). This increases the contribution of motional mode error by $1.5\times$ as the trap capacity increases to 35 ions. Second, heating of a long ion chain causes a large motional energy hot spot, worsening all gates in that trap. With small trap capacities, heating effects can effectively be localized to small regions of the device.

\emph{Therefore, for maximizing the reliability of QCCD systems, there is a trap capacity sweet spot of 15-25 ions, depending on the application. This capacity minimizes the impact of heating from communication, thermal motion of the laser beams and large hot spots on the device. Moreover, this trap sizing also offers very good runtime performance across applications.} 

TI devices can be easily reconfigured to support less ions than the trap maximum capacity, simply by loading fewer ions. Hence, we recommend that QCCD system should be designed to support up to 20-25 ions per trap. The actual used capacity can be lowered for applications which benefit from small trap sizes.

\subsection{Communication Topology Choices}
Figure \ref{fig:comm_choices} compares the execution time and fidelity of linear (L6) and grid (G2X3) communication topologies across applications. 

For Adder, QFT, Supremacy and QAOA the linear topology offers slightly better performance than grid. For SquareRoot, the grid topology offers better performance than linear. Comparing QFT and SquareRoot, SquareRoot has fewer two-qubit operations than QFT, but its communication pattern is more irregular. QFT has a very regular communication pattern where every ion communicates with every other ion in sequence. Hence, QFT maps well onto the linear topology and SquareRoot maps well onto the grid topology. Therefore, for a given architecture, application gate patterns significantly influence runtime performance.  

Comparing fidelities, topology has a dramatic impact on the fidelity of SquareRoot and QFT. For SquareRoot, the grid topology offers up to $7000\times$ higher fidelity than the linear topology. For QFT, the linear topology offers up to $4\times$ higher fidelity than grid. Figure \ref{fig:comm_choices}g shows the motional mode energies for SquareRoot. The grid topology offers benefits for SquareRoot because it reduces the number of split and merge operations at intermediate traps and therefore accrues less motional heating. The grid topology also allows shorter shuttling paths for the irregular communication pattern of this application, further minimizing unwanted motional energy. For Adder, BV, Supremacy and QAOA the impact of topology is less because they are not communication-intensive. In particular, Supremacy and QAOA (we use the hardware-efficient ansatz) are designed for nearest-neighbor connectivity and work well on QCCD systems with linear topology. 

\emph{Thus, device topology must be co-designed for needs of applications. For NISQ systems, fidelity losses from application-device topology mismatch can be very severe. For nearest-neighbor applications such as QAOA and Supremacy, linear QCCD topologies work well.}

\section{Microarchitectural Design Exploration}
\begin{figure*}[t]
    \centering
    \label{fig:gate_impln_choices}
    \includegraphics[trim = 0 0 0 0, width = \linewidth]{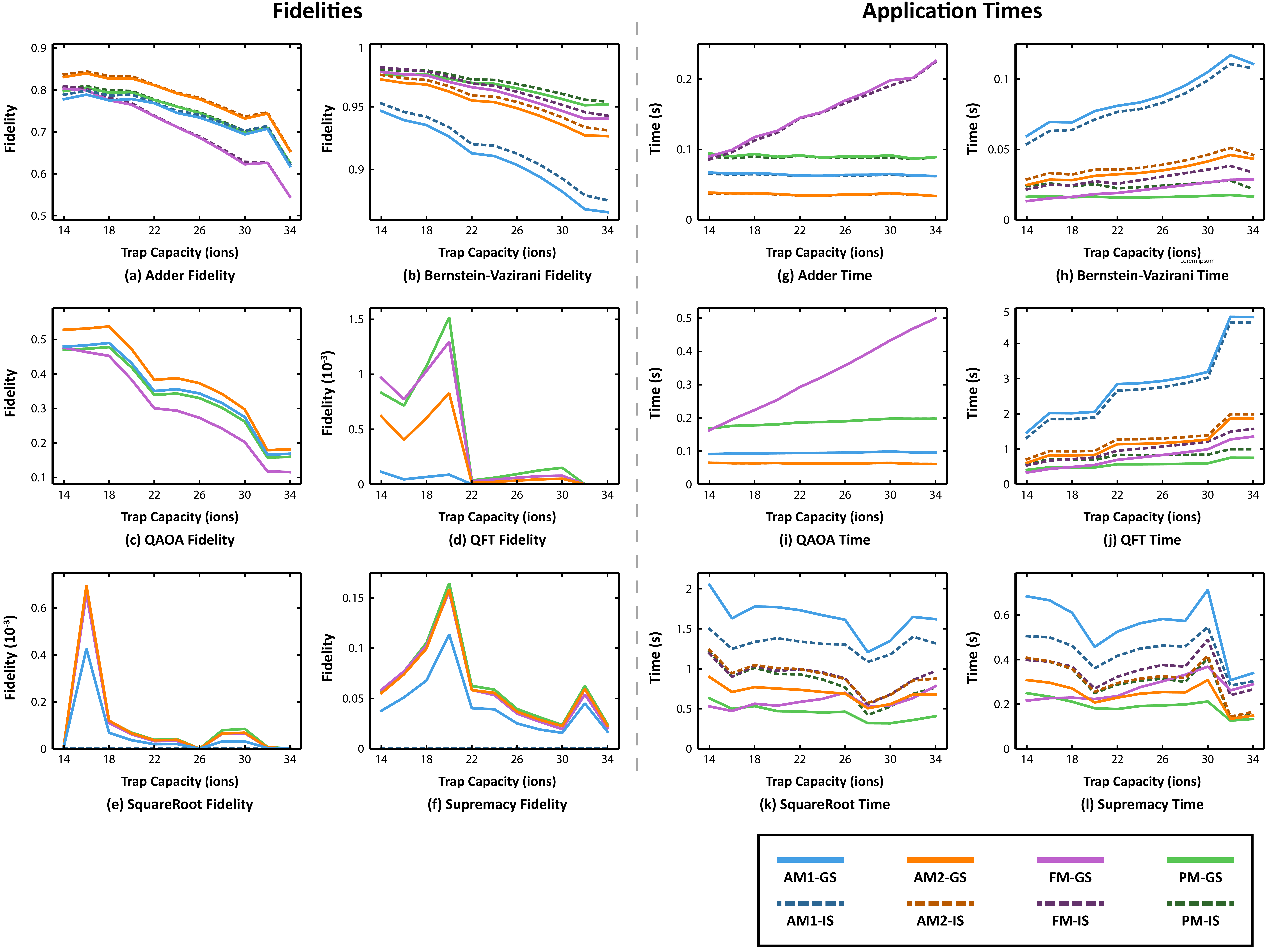}
\caption{\textbf{Effect of microarchitecture choices on application fidelity and performance.} Experiments use L6 topology. Comparison of 8 combinations with 4 gate choices: \amone, \amtwo, \pmgate, and \fm, and two chain reordering methods: \gs, and \is. Application fidelity varies across gate implementations. FM and PM have no or weak distance dependence respectively, so they work well for QFT and SquareRoot which requires several long-range gates. \pmgate or \amtwo\ gates work well for other applications because they offer fast and high-fidelity short-range interactions. Across applications, gate-based swapping is superior to physical swapping of ions in terms of fidelity because the latter requires many more split and merge operations.}
\label{fig:gate_impln_choices}
\end{figure*}
Figure \ref{fig:gate_impln_choices} shows the performance and fidelity for the six applications under eight microarchitecture combinations: four two-qubit gate implementation methods (\amone, \amtwo, \pmgate, \fm) and two chain reordering methods (\gs, \is). For this simulation, we used a linear device topology with 6 trapping zones. 

\subsection{Two-qubit Gate Choice}
The right half of Figure \ref{fig:gate_impln_choices} show the performance of the gate implementations. Application performance depends on the gate implementation, with up to $5\times$ performance variations across implementations. Thus best choice of gate differs according to the application. For QAOA where all the two-qubit are short range, AM gates perform better than the FM gate. This is because FM gates have high execution times which increase linearly with the number of ions in the chain. However, FM gate time is independent of the ion separation for a particular two-qubit gate and PM gates only have a weak distance dependence, and therefore they are suitable for SquareRoot and QFT which have long range two-qubit operations.

The left side of Figure \ref{fig:gate_impln_choices} show that application fidelity depends significantly on the two-qubit gate implementation choices (with the GS chain reordering method). Fidelity varies by up to $7\times$ across implementations (not including IS). FM gates obtain up to $9\times$ improvement over AM1 and $1.5\times$ over AM2. In most cases, FM and PM gates have comparable fidelities. For SquareRoot, FM and AM2 gates are comparable and obtain up to $2\times$ improvement over AM1. For QAOA, AM2 gates work well, for Supremacy FM, AM2 and PM gates are comparable and better than AM1. Similar to the performance variations, fidelity varies due to different application requirements. QAOA, Supremacy and Adder benefit from fast gate times at short-range, hence AM2 gates work well. QFT, SquareRoot, BV have short and long-range interactions which are reliably provided by the FM or PM implementations.

\emph{Therefore, QCCD systems should support multiple implementations for two-qubit gates to allow applications to matched to the most suitable implementation. The right choice of gate can improve fidelity by up to $9\times$. However, this will not require extra hardware --- current TI systems already include all the hardware necessary to allow experiments with different gate implementations \cite{blumel2019power}.}


\subsection{Chain reconfiguration Choices} 
Figure \ref{fig:gate_impln_choices} shows that GS chain reordering has superior fidelity to IS (for QAOA, GS and IS curves match exactly because chain reordering is not required). Although fast methods have been developed for \is\ \cite{ionswapping}, our simulations indicate that this method has severe fidelity overheads. With current protocols for reordering, each pair of adjacent ions requires an additional split and merge operation. Applications such as SquareRoot require several reordering operations, especially at small trap sizes, increasing the overheads of \is. \gs\ works well across applications, across FM and AM2 gates, and across different trap sizes, providing vastly superior fidelity compared to \is.

\emph{Thus, we recommend that QCCD-based TI systems used gate-based swapping for chain reordering. This method also has the advantage that it can leverage one or more two-qubit gate implementations available for the trap.} 

%


\section{Related Work}
Several works have developed software tools and architecture for superconducting systems. \cite{koen_bertels1, koen_bertels2, koen_bertels3, qumasim} developed the QuMA microarchitecture and simulation tools. \cite{1904.11590, alimicro50, tannu_qureshi, asplos} are other recent simulation and compilation works on superconducting systems. IBM's Qiskit \cite{qiskit}, Google's Cirq \cite{cirq} and Rigetti's Quilc \cite{quilc} platforms are software toolflows for their respective superconducting devices. Overall, superconducting systems have received significant attention recently, in part because several industry players have provided access to real devices. Our work brings renewed attention to TI systems which may offer comparable or higher reliability \cite{linke2017experimental, triq}.

Prior works have evaluated real implementations of TI systems to understand architecture design choices. \cite{linke2017experimental} compared the performance of a 5-qubit TI system from University of Maryland (UMD) with a 5-qubit superconducting system from IBM.\cite{triq} conducted a larger study, comparing superconducting systems from IBM and Rigetti to the UMD 5-qubit TI system. From \cite{linke2017experimental, triq}, the full connectivity of TI systems and powerful primitive operations offer high application success rates compared to other platforms. These real-system studies provide ample evidence to show that TI systems are very promising for NISQ applications. While these works focus on existing devices with less than 20 qubits, our work focuses on the 50-100 qubit range using an automated design and simulation toolflow.

Prior works have also considered very large or fault-tolerant TI devices. \cite{ahsan1, ahsan2, ahsan3} developed simulation tools for systems with million qubits. \cite{musiqc1, musiqc2} developed the MUSIQC architecture which uses optical interconnects to scale to thousands of qubits, \cite{saqip} proposed a scalable architecture for TI systems based on a reconfigurable optical interconnect. All these works focus on very large or fault tolerant systems which are unlikely to be realized in the next ten years \cite{nasemreport}. 

\section{Conclusions}
Current TI systems use a single-trap architecture where all qubits reside in the same ion chain. Realizing the scaling issues of this design, the Quantum Charge Coupled Device (QCCD) architecture was proposed \cite{qccd1} as a path towards modular TI systems. Over the last two decades, all components required for QCCD systems have been experimentally developed and honed. Recently, Honeywell demonstrated the first QCCD system with two traps and four qubits. However, building a practically useful QCCD system is challenging due to the wide range of possible hardware choices and the need to support an evolving mix of QC applications. While performance trends are known in isolation for individual components, there is little guidance on the their system-level performance or impact on applications.

In classical computing, architectural simulations have been a key enabler of technology progress, allowing us to predict the performance of the next generation of machines before building them. Our work proposes the use of computer architecture and simulation techniques to scale TI quantum systems to the next major milestone of 50-100 qubits. We build a design toolflow for the QCCD architecture, including an optimizing compiler and simulator. Using real performance models and device characterization data as inputs to our toolflow, we evaluate application runtime and reliability across several design possibilities. As a result, we provide design insights and recommendations for choosing trap sizes, topology, and gate implementations to maximize application reliability and performance. 
With several efforts underway to build large QCCD systems, our work has the potential to guide QC hardware design in the near future.




\section*{Acknowledgment}
DD would like to thank Pak Hong Leung and Ye Wang for helpful discussions regarding ion trap gates and errors. This work is funded in part by  Enabling Practical-scale Quantum Computation (EPiQC), an NSF Expedition in Computing, grants 1730082, 1730104 and Software-Tailored Architecture for Quantum co-design (STAQ) under NSF grant 1717523. 
\bibliographystyle{IEEEtran}


\end{document}